# Measurement of accelerator neutron radiation field spectrum by Extended Range Neutron Multisphere Spectrometers and unfolding program

LI Guanjia(李冠稼), WANG Qingbin(王庆斌), MA Zhongjian(马忠剑), GUO Siming(郭思明), YAN mingyang(阎明洋),SHI haoyu(石潓玙),XU chao(徐超)

Institute of High Energy Physics, CAS, Beijing, 100049, China

**Abstract:** This paper described a measurement of accelerator neutron radiation field at a transport beam line of Beijing-TBF. The experiment place was be selected around a Faraday Cup with a graphite target impacted by electron beam at 2.5GeV. First of all, we simulated the neutron radiation experiment by FLUKA. Secondly, we chose six appropriate ERNMS according to their neutron fluence response function to measure the neutron count rate. Then the U_M_G package program was be utilized to unfolding experiment data. Finally, we drew a comparison between the unfolding with the simulation spectrum and made an analysis about the result.

## 1    Introduction

Neutrons play a dominant role in the radiation field of large particle accelerate facilities, and the neutron spectrum energy distribute from thermal neutron (approximately 0.0253eV) up to very high energy neutron (for several GeV).Therefore, the option of instruments is significant to measure specific neutron field.

In the 1960s, Bramblett, Ewing and Bonner [1] invented a series of neutron detectors, which were designed as several $^3$He counter tube covered by polyethylene moderating spheres in different diameter (from 2 to 12 inch).Those detectors were called Bonner Sphere Spectrometer (BSS).As a typical neutron field measure instrument, the BSS is provided with extended sensitive energy range up to 20MeV. By attaching high-Z material (such as lead), the higher energy neutron could be moderated and detected by the $^3$He counter tube in the center. The improved BSS systems were called Extended Range Neutron Multisphere Spectrometers (ERNMS), which is our best choice to measure the neutron field spectrum in accelerate facilities.

In July 2015,an experiment to measure specific neutron spectrum was conducted at the Transport Lines of the Beijing-TBF (E2 line), which is part of the Beijing Electron-Positron Collider, in Institute of High Energy Physics. Six different size and designed sphere detectors were utilized in this experiment.

This paper provides a particular overview of this experiment, including the experiment environment and simulation, the option of the detectors, the compare with the results and data analysis.

## 2    Experimental environments and detectors

### 2.1    overview of the experimental place

The Linear Accelerator (LINAC), in the Institute of High Energy Physics, is constructed to provide high energy electron beam for the Beijing Electron-Positron Collider. The highest beam energy could get up to 2.5GeV. The experiment place was chosen at Transport Lines of the Beijing-TBF, E2 line, an outlet beam line from the LINAC, as shown in Fig 1.

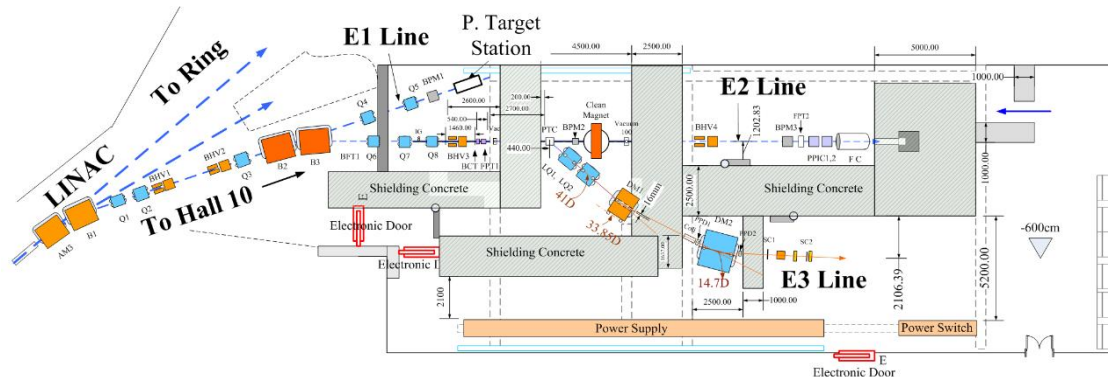

Fig. 1.   The configuration of Beijing-TBF on LINAC at Hall 10 of BEPCII Area

In this experiment, 2.5GeV energy electron beam was used to impact a graphic target inside a Faraday Cup to generate neutron, the electron number reached to $1.5*10^9$ particle per second, and the structure is shown in Fig 2.

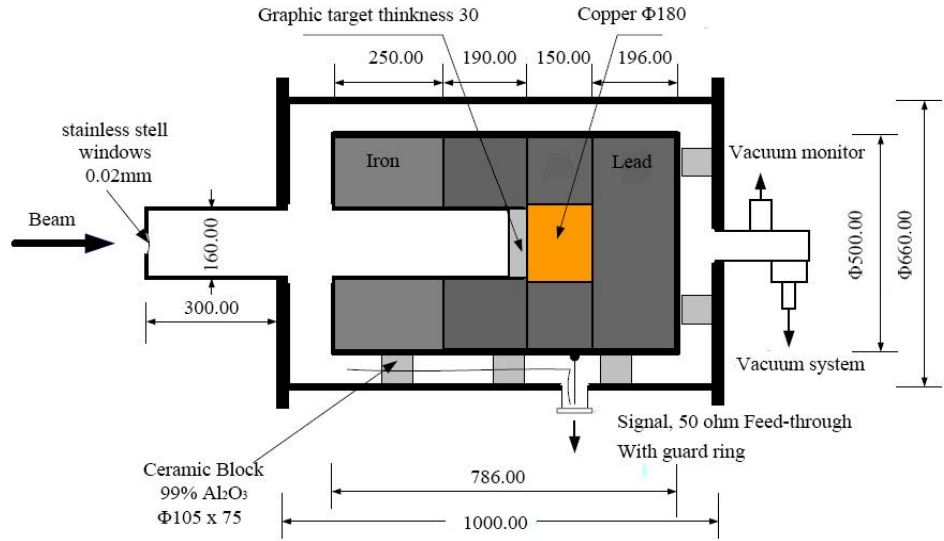

Fig. 2.   The design structure of Faraday Cup at E2 line

To simulate the condition of the radiation field, the option of simulation instrument became very significant for this experiment. FLUKA code [2] (version 2011.2c.0) played an important role to build the detailed geometry model, simulated the electron beam and gathered statistics the particle energy and number. According to USRBIN card provided by FLUKA, the neutron dose distribution of the Faraday Cup area could be evaluated. The chromatic graph of neutron dose distribution plotted by SimpleGeo 4.3 [3] is shown in Fig 3.

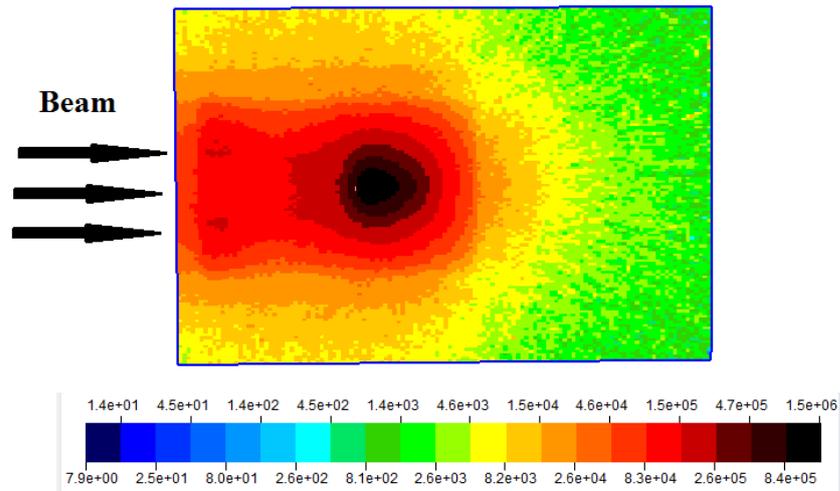

Fig. 3.   The chromatic graph of neutron distribution and the figure legend

2.2   the detectors case

The experiment was meant to measure the neutron dose rate of an area beside the Faraday Cup, then utilized unfolding programs to achieve the neutron field spectrum and compared with the simulation spectrum. In terms of a radiation field, the neutron distribute from thermal neutron energy (0.0253eV) to very high energy (>10GeV).Conventional Bonner Sphere Spectrometers could not cover all of that energy. In consequence the Extended Range Bonner Sphere Spectrometers were the best choice to gain data from radiation field because of the large sensitive neutron energy range.

Normally the measuring result precision of the neutron spectrum depends on the counts of detectors, which means more different size of detectors should be used. However, due to the limitation of experiment place, so many detectors would produce deviation when measuring the same location. Considering the situation of the neutron field around the Faraday Cup, six different size of ERNMS [4] were chosen.

The ERNMS system consisted of one bare $^3$He proportional counter, four different diameters Bonner detectors and one Bonner detector with lead layer inside. These four Bonner Spheres were designed in separately 4inch, 6inch, 8inch and 12inch diameter polyethylene sphere covered their $^3$He counter. The Bonner detector with lead was a 3inch diameter polyethylene sphere covered by a 5inch diameter lead spherical shell, and then covered with a 6inch diameter polyethylene spherical layer (the 3P5L6P sphere). Figure 4 shows the neutron fluence response function of those six detectors calculated by FLUKA.

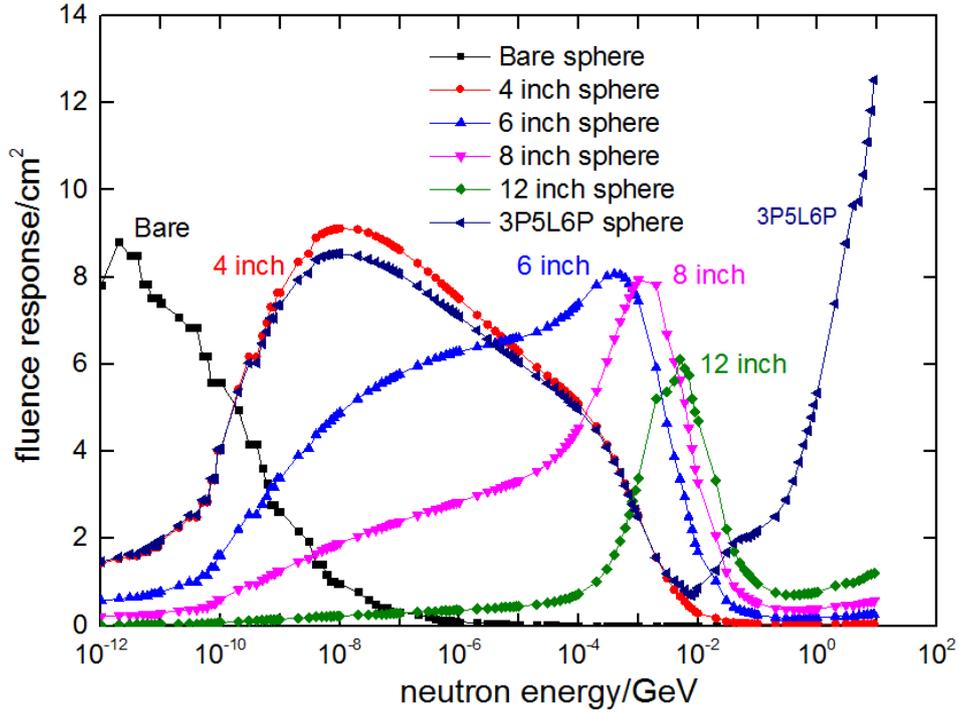

Fig. 4.  The neutron fluence response function of neutron energy calculated by FLUKA, for the six different diameters spheres in this experiment.

## 3  Measurements

The measurement location was selected 1.5 meters from the Faraday Cup's sideway. The counters inside the spheres would transport signal to amplifying circuit and filter, then the SCMs could receive the digital signal and reserve. All of the detectors were connected with the switch to transport data to a personal computer, and the data collection system would extract data from every SCM per five minutes. By this means neutron count rate per second of every detectors could be converted.

For this experiment the whole measuring system had been operated for 10 hours, the conversion neutron rate and error value were expressed in Table 1.

Table 1.  Neutron count rate and error value of six sphere detectors.

| Detector type | Neutron count rate (Particle/second) | Error value (Particle/second) | Relative value (%) |
|---|---|---|---|
| Bare sphere | 105.76 | 1.64 | 1.55 |
| 4 inch sphere | 96.76 | 1.60 | 1.65 |
| 6 inch sphere | 163.85 | 2.67 | 1.63 |
| 8 inch sphere | 50.86 | 0.43 | 0.85 |
| 12 inch sphere | 154.13 | 6.46 | 4.19 |
| 3P5L6P sphere | 74.31 | 0.50 | 0.67 |

According to Tab.1, all of the relative value of the measurements were under 5%, which manifested that the ERNMS measuring systems were stabilized, and the results were available. On the basis of the neutron count rate, the spectrum could be calculated by the unfolding methods.

## 4  Unfolding methods

To achieve the neutron spectrum, an effective unfolding method was essential in this experiment, while the U_M_G [5] (unfolding with Maxed and Gravel) package met the demand. The U_M_G package included MXD_FC31, GRV_FC31, and IQU_FC31, which was prepared by employees of Physikalisch-Technische Bundesanstalt (PTB), was written specifically for the unfolding of Bonner sphere data.

### 4.1  unfolding algorithm principle

The essence of spectrum unfolding was to find the most reliable spectrum from the whole probable spectrums. Normally the spectrum unfolding principle could be expressed as the following equations:

$$N_k + \varepsilon_k = \int R_k(E)\Phi(E)\,dE \quad k=1, 2, \ldots n \qquad (1)$$

Where $n$ = the number of detectors, $N_k$ = the measured counts by number $k$ detector, $\varepsilon_k$ = measure uncertainty by number $k$ detector, $R_k(E)$ = the fluence response when neutron energy is $E$ by number k detector, $\Phi(E)$ = the fluence when neutron energy is $E$ by number k detector. In particle, the energy was not continuous as the integral form but discrete, and the energy would be divided in several bins. So previous equation could be written as following equations:

$$N_k + \varepsilon_k = \sum_i R_{ki}\Phi_{ki} \quad k=1, 2, \ldots n\,;\quad i=1, 2, \ldots m \qquad (2)$$

Where the $m$ = the number of energy divided bins, $R_{ki}$ = the fluence response when neutron energy is in number $i$ energy bins area, $\Phi_{ki}$ = the fluence when neutron energy is in number $i$ energy bins area.

Once the measure counts and the response fluence have been known, the spectrum could be calculated by deconvolution method in terms of equation (2). When n<m, in other words, the number of detectors was less than the number divided in energy bins, was called "few-channel" unfolding methods. U_M_G belonged to this methods.

### 4.2  MXD_FC31

The MXD_FC31 was derived from the MAXED (MAXimum Entropy Deconvolution) program, and improved on the unfolding algorithm. MAXED program centered on maximum entropy principle. To achieve the spectrum a default spectrum that attained from experience or calculated was necessary. The essence of unfolding was to find a spectrum that closet to the default from the whole spectrums matched the measurement condition. In order to get "the closest one", a relative entropy $S$ was involved shown as follow:

$$S = \sum_i [\Phi_i \ln(\Phi_i/\Phi_i^{DEF}) + \Phi_i^{DEF} - \Phi_i] \tag{3}$$

According to maximum entropy principle, the unfolding spectrum should allow relative entropy $S$ maximum. When solving the problem, two limitation should be considered, which were written in formula (2) and formula (4) shown as follow:

$$\sum_k \frac{\varepsilon_k^2}{\sigma_k^2} = \chi^2 \tag{4}$$

Where $\Phi_i^{DEF}$= default spectrum fluence, $\sigma_k$= standard error of the measurements, $\chi^2$=parameter according with chi square distribution, normally equals to 1 or the number of detectors.

MXD_FC31 had an improvement on the basis of MAXED, which involved a new limitation to calculate the maximum entropy shown as formula (5):

$$\sum_k \frac{N_k}{\sigma_k} - \sum_{k,i} \frac{R_{ki}\Phi_i}{\sigma_k} = 0 \tag{5}$$

The output file of MXD_FC31 could be regarded as the input file of the IQU_FC31 program to statistic relevant integration variable and analysis the uncertainty.

4.3  GRV_FC31

The GRV_FC31 was a different unfolding program involved in U_M_G package, and improved from the SAND-II program. SAND-II program was devised in the 1970s, as the most general unfolding methods. The iteration formula was shown as equation (6):

$$\Phi_i^{J+1} = \Phi_i^J \exp\left[\frac{\sum_k W_{ik}^J \log\left(\frac{N_k}{\sum_{i'} R_{ki'}\Phi_{i'}^J}\right)}{\sum_k W_{ik}^J}\right] \tag{6}$$

Where

$$W_{ik}^J = \frac{R_{ki}\Phi_i^J}{\sum_{i'} R_{ki'}\Phi_{i'}^J} \frac{N_k^2}{\sigma_k^2} \tag{7}$$

At the process of unfolding, a non-negative default spectrum was necessary as the first order iteration spectrum, that was to say when $k$=1, $\Phi_i^1 = \Phi_i^{DEF}$. Both the time of iteration and the $\chi^2$ parameters should be set, and the program would stop iterating either times of iteration or $\chi^2$ reach the set value.

The output file of GRV_FC31 could also be regarded as the input file of IQU_FC31 program but it could only statistic the relevant integration variable not the uncertainty

## 5  Result and discussion

### 5.1  the simulation neutron spectrum

As the preceding explanation said, a default spectrum was essential to the unfolding program. The spectrum should be close to the real neutron field so that the unfolding result could be available. FLUKA provided a scoring card called **USRTRACK** which based on track-length. The **USRTRACK** card needed the estimation of volume-averaged fluence (differential in energy) for any particle in any selected region and then calculated the track length of the specific particle in the region. The unit of track-length was cm/particle, and the total response could be calculated according to divide the track-length by the volume of detector area (in one particle/cm$^2$ /particle). After two binning estimator programs, **ustsuw.f** and **usbrea,f**, a table of energy bins and particle fluence for each energy intervals was outputted. According to the table, the neutron spectrum could be plotted, and it would be regarded as the default spectrum of unfolding.

### 5.2  the unfolding of the spectrum

By the use of default spectrum, the unfolding programs could work out the experimental neutron spectrum according to the two methods, MXD_FC31 and GRV_FC31. Both of the chi square parameter $\chi^2$ was selected in 1.8, and the default spectrum was normalized and processed after simulation. The comparison graph between the default spectrum and the unfolding results was shown as Figure 5.

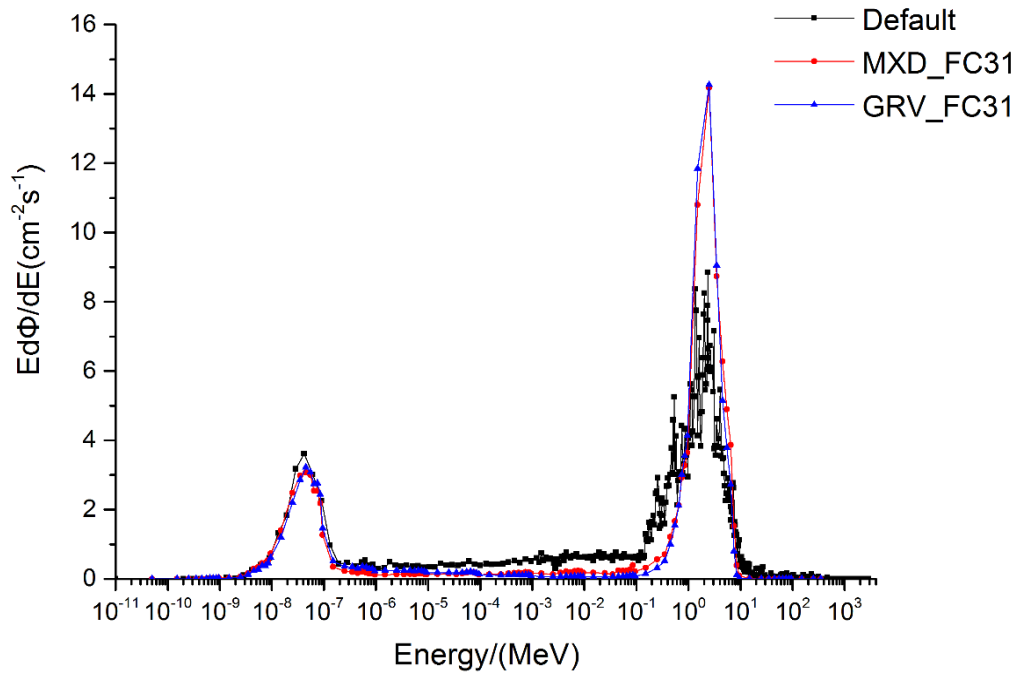

Fig. 5.   The unfolding spectrum result by MXD_FC31 and GRV_FC31, and contracted with the default spectrum which was normalized and processed.

The figure above showed that the unfolding results accord well to the simulation in the thermal neutron energy region, while in the intermediate region from 10eV to 1MeV the according degree became rough; in the region from 1MeV to 10MeV the fluence peak emerged while the unfolding ones were higher than the simulation; the fluence vanished gradually when the energy accessed 10MeV [6]. The unfolding statistic relevant integration variable was shown in Table 2, and Table 3 showed the percentage of each integration parts.

Table 2. The unfolding statistic relevant integration variable, the total integration was divided in four parts, the thermal neutron region (E < 0.5eV), the intermediate region (0.5eV<E<0.1MeV), the fast neutron region (0.1MeV<E<10MeV) and the very fast neutron region (E>10MeV)

| Unfolding program | $\varphi_{total}$ (cm$^{-2}$s$^{-1}$) | $\varphi_{thermal}$ (cm$^{-2}$s$^{-1}$) | $\varphi_{inter}$ (cm$^{-2}$s$^{-1}$) | $\varphi_{fast}$ (cm$^{-2}$s$^{-1}$) | $\varphi_{from10}$ (cm$^{-2}$s$^{-1}$) |
|---|---|---|---|---|---|
| MXD_FC31 | 32.35 | 6.96 | 2.28 | 23.03 | 0.05 |
| GRV_FC31 | 32.03 | 6.99 | 1.99 | 23.01 | 0.03 |

Table 3. The percentage of each integration parts which divided in Table 2

| Unfolding program | $\varphi_{total}$ (%) | $\varphi_{thermal}$ (%) | $\varphi_{inter}$ (%) | $\varphi_{fast}$ (%) | $\varphi_{from10}$ (%) |
|---|---|---|---|---|---|
| MXD_FC31 | 100 | 21.51 | 7.05 | 71.19 | 0.01 |
| GRV_FC31 | 100 | 21.82 | 6.21 | 71.84 | 0.01 |

5.3 the discussion of the result

As the integrations percentage showed, thermal and fast neutrons were dominating the neutron field spectrum, and the fluence was focused on the evaporation component [7] (0.1MeV to 10MeV), which caused by the quasi-elastic interaction of particle-nucleus. The electron impacted the graphite target and liberated various energy neutron. By the moderate materials such as concrete and lead the energy declined to thermal energy region. In fact the moderate efficiency was limited at the experiment place, most of the neutron interacted with nucleus by inelastic scattering and deposited energy at 0.1MeV to 10MeV energy region.

From the comparison between simulation and unfolding, more fast neutron was detected than the simulation. As the simulation condition was idealized than the reality, some details of the layout could not be reflected in the FLUKA input file. On the other hand, the limitation of the detectors number also affected the unfolding result. More different size of detectors could provide larger range of neutron fluence response, and the detector data would be more available. While using more detectors means to occupy broader placement area. The situation would lead to bias both on simulation and measurement [8].

From above all, the experiment reflected the authentic spectrum of the neutron field, and the result could be a reference to the protection of accelerator radiation.

## 6    Conclusion

The experiment was aimed at measure the neutron field spectrum of E2 beam line out from LINAC. Firstly, we simulated the Faraday Cup's situation to achieve the spectrum by FLUKA. Secondly, by the use of six ERNMS detectors we got the neutron count rate in the specific area. Then we used U_M_G package program to unfold the data and attain the experiment spectrum. Finally we compared the simulation spectrum with the unfolding ones and made an analysis of the result. The essential difference was mainly because of idealization of simulation, layout of detectors and the unfolding methods. In a word, using of ERNMS to measure the accelerator neutron field is quite valid, to make the result more reliable we should make sure the simulation more according with the reality, choose more suitable detectors and improve the unfolding methods.

# 扩展型多球中子谱仪及解谱程序对加速器辐射中子场的测量

李冠稼[*],王庆斌,马忠剑,郭思明,阎明洋,石澔玙,徐超

中国科学院高能物理研究所,北京,100049,中国

**摘要**:本文旨在介绍北京正负电子对撞机直线加速器引出束流的中子辐射场能谱测量,实验地点选择在 2.5GeV 电子束流撞击法拉第杯石墨靶处。首先对实验环境进行 FLUKA 模拟计算,其次根据扩展型多球中子谱仪能量通量响应选择六个探测器来进行测量中子计数率。接着利用 U_M_G 解谱软件包对测量数据进行解谱,最后将解谱结果与模拟得到的能谱进行对比和分析。

**关键词**:中子辐射场能谱、扩展型多球中子谱仪、能量通量响应、FLUKA、解谱方法